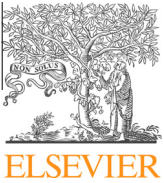
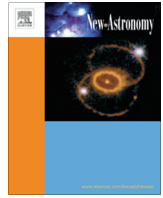

# Photometric observation of HAT-P-16b in the near-UV

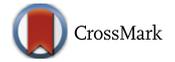

Kyle A. Pearson [a,*], Jake D. Turner [b], Thomas G. Sagan [a]

[a] Steward Observatory, University of Arizona, Tucson, AZ 85721, USA
[b] Lunar and Planetary Laboratory, University of Arizona, Tucson, AZ 85721, USA

## HIGHLIGHTS

- We present the first near-UV light curve of HAT-P-16b.
- The magnetic field of HAT-P-16b has been constrained.
- ExoDRPL, Exoplanet Automated Data Reduction pipeline, was developed.
- EXOMOP, Exoplanet Modeling Package, was also developed for this project.

## ABSTRACT

We present the first primary transit light curve of the hot Jupiter HAT-P-16b in the near-UV photometric band. We observed this object on December 29, 2012 in order to update the transit ephemeris, constrain its planetary parameters and search for magnetic field interference. Vidotto et al. (2011a) postulate that the magnetic field of HAT-P-16b can be constrained if its near-UV light curve shows an early ingress compared to its optical light curve, while its egress remains unchanged. However, we did not detect an early ingress in our night of observing when using a cadence of 60 seconds and an average photometric precision of 2.26 mmag. We find a near-UV planetary radius of $R_p = 1.274 \pm 0.057 R_{Jup}$ which is consistent with its near-IR radius of $R_p = 1.289 \pm 0.066 R_{Jup}$ (Buchhave et al., 2010). We developed an automated reduction pipeline and a modeling package to process our data. The data reduction package synthesizes a set of IRAF scripts to calibrate images and perform aperture photometry. The modeling package utilizes the Levenberg–Marquardt minimization algorithm to find a least-squares best fit and a differential evolution Markov Chain Monte Carlo algorithm to find the best fit to the light curve. To constrain the red noise in both fitting models we use the residual permutation (rosary bead) method and time-averaging method.

Published by Elsevier B.V.

## 1. Introduction

HAT-P-16b, one of ~317[1] confirmed transiting exoplanets, is a hot Jupiter orbiting a F8-type star with a short orbital period of $2.77596 \pm 0.000003$ d, a mass of $4.193 \pm 0.094\ M_{Jup}$, and a radius of $1.289 \pm 0.066\ R_{Jup}$ (Buchhave et al., 2010). Several follow-up spectroscopic studies have constrained additional stellar and planetary parameters, Husnoo et al. (2012) confirmed that the planet had the smallest reliably measured eccentricity of $0.034 \pm 0.003$. There have been no additional photometric measurements or secondary transit measurements of this planet yet.

The transit method for detecting and observing exoplanets is the only method that allows direct measurements of the planetary radius and can also be used to determine planetary parameters such as surface gravity, average density, atmospheric composition, semi-major axis and eccentricity (Charbonneau et al., 2007). We can obtain information about certain absorption features of the planet as light gets filtered through a planet's atmosphere during the primary transit. Probing the optical and near-UV wavelengths where the opacity is dominated by Rayleigh scattering can give us insight to information about clouds at high-altitudes (Pont et al., 2008), atmospheric circulation patterns (Knutson et al., 2007) and possible interior structures and compositions (Charbonneau et al., 2002). In addition, Lazio et al. (2010a) theorized that the transit method can be used to detect exoplanet magnetic fields.

The gas giant planets in our Solar System possess magnetic fields (Russell and Dougherty, 2010) and it has been predicted through interior structure models that extrasolar gas giants should have magnetic fields (Sánchez-Lavega, 2004). Detecting and studying the magnetic fields of exoplanets will allow for the investigation of more properties of exoplanets, including interior structure and rotation period (Lazio et al., 2010a), atmospheric retention (Lazio et al., 2010a; Cohen et al., 2011), and the presence of extrasolar moons (e.g., the variability in Jupiter's magnetic field can be attributed to the presence of Io (Lazio et al., 2010a)). Griemeier et al. (2005) suggest that the magnetic field of Earth helps contribute to its habitability by deflecting cosmic rays and stellar wind particles; exoplanets could also exhibit this characteristic (Lazio et al., 2010a). Studying the magnetic fields of hot Jupiters will help lay the foundation for the characterization of magnetic fields around Earth-like and other types of planets, and therefore, it will aid in the search for life outside our Solar System.

---

* Corresponding author.
 E-mail addresses: pearsonk@email.arizona.edu (K.A. Pearson), turner@email.arizona.edu (J.D. Turner), sagant@email.arizona.edu (T.G. Sagan).
 [1] http://exoplanetapp.com/.

There are several methods that can be used in attempts to detect magnetic fields of exoplanets. Farrell et al. (1999), Zarka et al. (2001), and Lazio et al. (2010b) suggest that the most direct method for detecting the magnetic field of an exoplanet is through radio emission from the planet generated by electron-cyclotron maser interactions. This electron-cyclotron maser radio emission is generated as new material is injected into the magnetosphere due to interactions between the solar wind and magnetosphere. Many studies conducted to find exoplanet radio emissions have resulted in non-detections (e.g., Yantis et al., 1977; Winglee et al., 1986; Ryabov et al., 2004; George and Stevens, 2007; Lazio and Farrell, 2007; Smith et al., 2009; Lecavelier Des Etangs et al., 2009; Lecavelier Des Etangs et al., 2011; Lazio et al., 2010b; Lazio et al., 2010c) and one possible detection by Lecavelier des Etangs et al. (2013) of the hot Neptune, HAT-P-11b. Alternatively, Cuntz et al. (2000), Saar and Cuntz (2001), and Ip et al. (2004) proposed that the interaction of an exoplanet's magnetic field with that of its host star could produce detectable changes in the star's outer layers and corona in phase with the planet's orbit. This indirect method of detecting the magnetic field of an exoplanet was validated through observations of Ca II H K emission by Shkolnik et al. (2003, 2005, 2008), and Gurdemir et al. (2012). There is also a non-detection by Miller et al. (2012) while observing WASP-18b. However, it is possible an observational bias can lead to spurious trends of star-planet-interaction signatures due to planet detection methods and should be taken into consideration (Poppenhaeger and Schmitt, 2011).

In this paper we use another method to attempt to detect interference in our light curve due to the magnetic field of an exoplanet, described by Vidotto et al. (2010, 2011a,b,c), Lai et al. (2010), and Llama et al. (2011). They proposed that if a transiting exoplanet harbors a magnetic field it will show an early transit ingress in the near-ultraviolet (near-UV) wavelengths but not at longer wavelengths, while the transit egress will remain the same. The authors explained this effect by the presence of a bow shock in front of the planet formed by the interactions between the stellar coronal material and the planet's magnetosphere. Moreover, if the shocked material is sufficiently optically thick, it will absorb starlight and cause an early ingress in the near-UV light curve (Vidotto et al., 2011b, see Fig. 6). An early near-UV ingress has been observed in one transiting exoplanet, WASP-12b (Fossati et al., 2010 hereafter FHF10). Observations by FHF10 of WASP-12b using the *Hubble Space Telescope* with the NUVA (253.9–258.0 nm) near-UV filter indicate that the near-UV transit started approximately 25–30 min earlier than its optical transit. Using these observations, Vidotto et al. (2010) determined an upper limit for the magnetic field of WASP-12b to be ~24 G.

We can constrain the properties of the planet's magnetic field by analyzing the difference in ingress times in different wavelength bands. However, VJH11a do not go into detail about whether this effect can only be seen in narrow-band spectroscopy or broad-band near-UV photometry. Additionally, Vidotto et al. (2011c) predicted that bow shock variations should be common and are caused by eccentric planetary orbits, azimuthal variations in coronal material (unless the planet is in the corotation radius of the star), and time-dependent stellar magnetic fields (e.g., coronal mass ejections, magnetic cycles, stellar wind changes). Consequently the near-UV light curve of exoplanets predicted by VJH11a will exhibit temporal variations.

We chose HAT-P-16b for our study because it is listed as one of the possible candidates predicted by VJH11a to exhibit near-UV asymmetries. In addition, the WASP-12 and HAT-P-16 systems have very different physical characteristics (see Table 1 for a summary). We also include information on TrES-3b since its magnetic field was previously constrained by Turner et al. (2013) to be between 0.013 to 1.3 G and HAT-P-11b has a possible upper limit on its magnetic field to be 50 G (Lecavelier des Etangs et al., 2013). Therefore, since FHF10 observed near-UV asymmetries in WASP-12b and all of these planets are eccentric perhaps a more diverse planet(ie. one with different properties), HAT-P-16b, could exhibit this effect.

If observed, a difference in timing between the near-UV and optical light curves of HAT-P-16b can be used to determine the planetary magnetic field, $B_p$, with the following equation derived from Vidotto et al. (2011b):

$$B_p = B_* \left(\frac{R_*}{aR_p}\right)^3 \left\{\frac{2\delta t}{t_d}\left[\left(R_*^2 - \left\{\frac{a\cos i}{R_*}\right\}^2\right)^{1/2} + R_p\right] + R_p\right\}^3, \quad (1)$$

where $B_*$ is the host star's magnetic field, $R_*$ is the host star's radius, $a$ is the semi-major axis, $R_p$ is the planet radius, $\delta t$ is the difference in timing between the near-UV and optical ingress (U and V bands respectively), $t_d$ is the optical transit duration, and $i$ is the orbital inclination. The parameters $R_p$, $a$, $t_d$, and $i$ can all be derived from the near-infrared light curve, and $B_*$ and $R_*$ have been determined by previous studies (e.g., Buchhave et al., 2010) of the HAT-P-16 host star. VJH11a predicts a minimum planetary magnetic field required to sustain a magnetosphere for HAT-P-16b to be $B_p^{\min} = 0.0027 B_*$. Using Eq. (1), we derived a minimum timing difference between the near-UV and optical ingress times of 0.0042 s.

**Table 1**
Comparison of the HAT-P-16, HAT-P-11, TrES-3 and WASP-12b systems.

| Planet name | $M_b$ ($M_{\text{Jup}}$) | $R_b$ ($R_{\text{Jup}}$) | $P_b$ (d) | $a$ (au) | Spec. type | $M_*$ ($M_\odot$) | $R_*$ ($R_\odot$) | [Fe/H] | $B_p/B_*^a$ (%) | $\delta t^b$ (s) |
|---|---|---|---|---|---|---|---|---|---|---|
| HAT-P-11b | 0.08 | 0.45 | 4.89 | 0.050 | K4 | 1.29 | 1.46 | 0.31 | 0.03 | 84 |
| HAT-P-16b | 4.19 | 1.29 | 2.77 | 0.041 | F8 | 1.22 | 1.24 | 0.17 | 0.27 | 0.0042 |
| TrES-3b | 1.92 | 1.31 | 1.31 | 0.023 | G | 0.90 | 0.80 | -0.19 | 0.47 | 3 |
| WASP-12b | 1.41 | 1.79 | 1.09 | 0.023 | G0 | 1.35 | 1.57 | 0.30 | 3.2 | 5 |

[1] Reference for $M_b$, $R_b$, $P_b$, $a$, Spec Type, $M_*$, $R_*$, [Fe/H]; Bakos et al. (2010).
[2] Reference for $M_b$, $R_b$, $P_b$, $a$, Spec Type, $M_*$, $R_*$, [Fe/H]; Buchhave et al. (2010).
[3] Reference for $M_b$, $R_b$, $P_b$, $a$, Spec Type, $M_*$, $R_*$, [Fe/H]; Sozzetti et al. (2009).
[4] Reference for $M_b$, $R_b$, $P_b$, $a$, Spec Type, $M_*$, $R_*$, [Fe/H]; Hebb et al. (2009).
[a] $B_p/B_*$ is the minimum planetary magnetic field relative to the stellar one that is required to sustain a magnetosphere. Reference for $B_p/B_*$; Vidotto et al. (2011a).
[b] $\delta t$ is the minimum timing difference between the optical and near-UV ingress times calculated from Eq. (1) inputting $B_p/B_*$. Reference for $\delta t$; Vidotto et al. (2011a).

**Table 2**
Journal of observation.

| Date (UT) | Filter[1] | Telescope[2] | Cadence | OoT RMS[3] (mmag) | Res RMS[4] (mmag) | White noise (mmag) | Red Noise[5] (mmag) | Seeing ($''$) |
|---|---|---|---|---|---|---|---|---|
| 2012 Dec 29 | U | Kuiper | 60.45 | 2.26 | 2.27 | 2.34 | 0.00 | 1.46–3.19 |

[1] Filter used: U is the Bessel U (303–417 nm).
[2] Telescope used: Kuiper is the 1.55-m Kuiper Telescope.
[3] Out-of-Transit (OoT) root-mean-squared (RMS) relative flux.
[4] Residual (res) RMS flux after subtracting the EXOMOP best-fitting model from the data.
[5] Red Noise (temporally-correlated noise) calculated from EXOMOP.

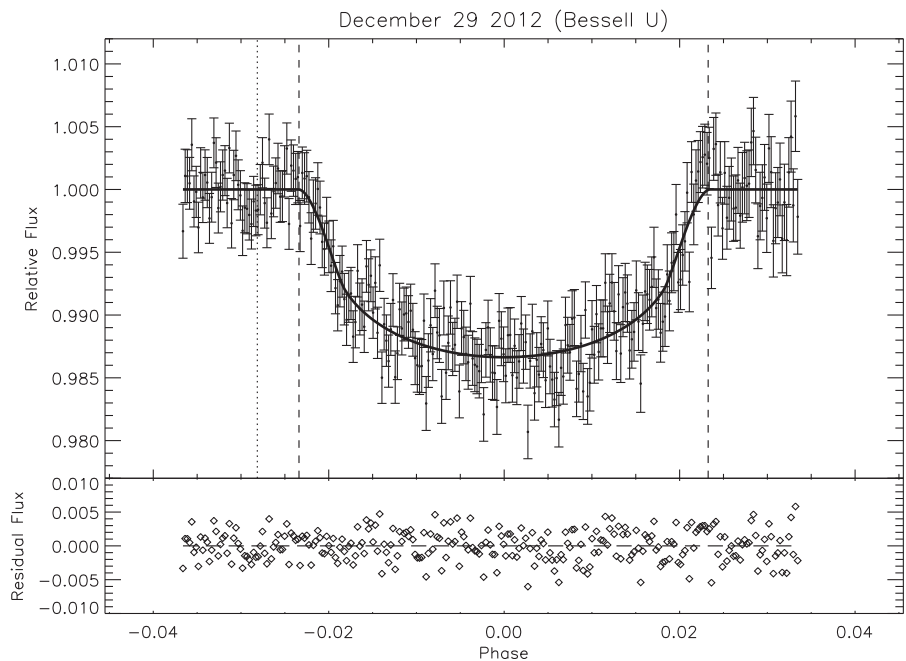

**Fig. 1.** Near-UV light curve of HAT-P-16b. The best-fitting model obtained from EXOMOP is shown as a solid black line. The residual is shown below the transit light curve. The EXOMOP best-fitting model predicted ingress and egress points are shown as dashed lines. The dotted line is the minimum timing difference (19 mins) between the near-UV and optical ingress found by using a reasonable estimate of $B_* = 100$ G and $B_p$ of 8 G (Reiners and Christensen, 2010). See Table 5 for the cadence, Out-of-Transit (OoT) root-mean-squared (RMS) flux, residual RMS flux, white and red (temporally-correlated) noise of the light curve.

The goal of this paper is to determine if our ground-based observation of the HAT-P-16b transit in broad-band near-UV wavelengths is capable of detecting the planet's magnetic field. Additionally, using our dataset, we update the planetary system parameters, the ephemeris and search for a wavelength dependence in the planetary radius.

## 2. Observations and data reduction

Our observation was conducted on December 29, 2012 at the Steward Observatory 1.55-m Kuiper Telescope on Mt. Bigelow, using the Mont4k CCD. The Mont4k CCD contains a $4096 \times 4096$ pixel sensor with a field of view (FOV) of 9.7 arcmin $\times$ 9.7 arcmin. We used a binning of $3 \times 3$ to achieve a resolution of $0.43''$/pixel which shortened our read out time to $\sim 15$ s. Our observation was taken with the near-UV filter, Bessell-U (303–417 nm), which has a transmission peak of 70% near 370 nm. To ensure accurate time keeping, an on-board system clock was automatically synchronized with a global positioning system every few seconds throughout the observational period. Due to excellent autoguiding, there was no more than a 2.5 pixel ($\sim 0.47''$) drift in the x position and 1.3 pixel ($\sim 0.99''$) drift in the y position of HATP-16-b. The seeing ranged from $1.46''$ to $3.19''$ throughout our observation. The Out-of-Transit (OoT) baseline in our transit achieved a photometric root-mean-squared (RMS) value of 2.27 millimagnitude (mmag), which is a typical value for high signal-to-noise (S/N) transit photometry on the 1.55-m Kuiper telescope (Dittmann et al., 2010, 2012; Scuderi et al., 2010; Turner et al., 2013; Teske et al., 2013). Our observation is summarized in Table 2.

To reduce the data we developed an automated reduction pipeline called ExoDRPL[2] that generates a series of IRAF[3] scripts that will calibrate images using the standard reduction procedure and perform aperture photometry. Each of our Kuiper 1.55 m images were bias-subtracted and flat-fielded. Turner et al. (2013) determined that using different numbers of flat-field images (flats) in the reduction of Kuiper Telescope/Mont4k data did not significantly reduce the noise in the resulting images. To optimize telescope time, we used 10 flats and 10 bias frames in our observation and reduction.

To produce the light curve we performed aperture photometry (using the task PHOT in the IRAF DAOPHOT package) by measuring the flux from our target star as well as the flux from eight different reference stars at 110 different aperture radii. For the analysis we used a constant sky annulus of 16 pixels, which was chosen to be a radius greater than the target aperture so that no stray light from the star was included. A synthetic light curve was produced by averaging the light curves from our reference stars, and the final light curve of HAT-P-16b was normalized by dividing by this synthetic light curve. Every combination of reference stars and aperture radii between five and sixteen at a precision of 0.1 were considered. We systematically chose the best reference stars and aperture by finding the lowest scatter in the OoT data points. The final light curve is shown in Fig. 1 with error bars of $1\sigma$ on each point derived from our IRAF reduction. The data points of our transit are available in electronic form (see Table 3).

---
[2] ExoDRPL is available from https://goo.gl/TLLZX.
[3] IRAF is distributed by the National Optical Astronomy Observatory, which is operated by the Association of Universities for Research in Astronomy, Inc., under cooperative agreement with the National Science Foundation.

**Table 3**
Photometry of HAT-P-16b[1].

| Time (BJD) | Relative Flux | Error Bars | X-pos of Star on CCD | Y-pos of Star on CCD | Airmass |
|---|---|---|---|---|---|
| 2456290.557216 | 0.9963810825 | 0.0021503996 | 795.674 | 719.398 | 1.016509 |
| 2456290.557915 | 1.0007790934 | 0.0021269843 | 795.161 | 719.303 | 1.016380 |
| 2456290.558614 | 1.0007598865 | 0.0021041695 | 795.084 | 719.531 | 1.016264 |
| 2456290.559313 | 1.0001904523 | 0.0020635430 | 795.573 | 719.540 | 1.016160 |
| 2456290.560011 | 1.0032490363 | 0.0020719485 | 795.655 | 719.121 | 1.016068 |
| 2456290.56071 | 0.9987371813 | 0.0020319225 | 795.862 | 718.939 | 1.015989 |

[1] This table is available in its entirety in machine-readable form in the on line journal. A portion is shown here for guidance regarding its form and content.

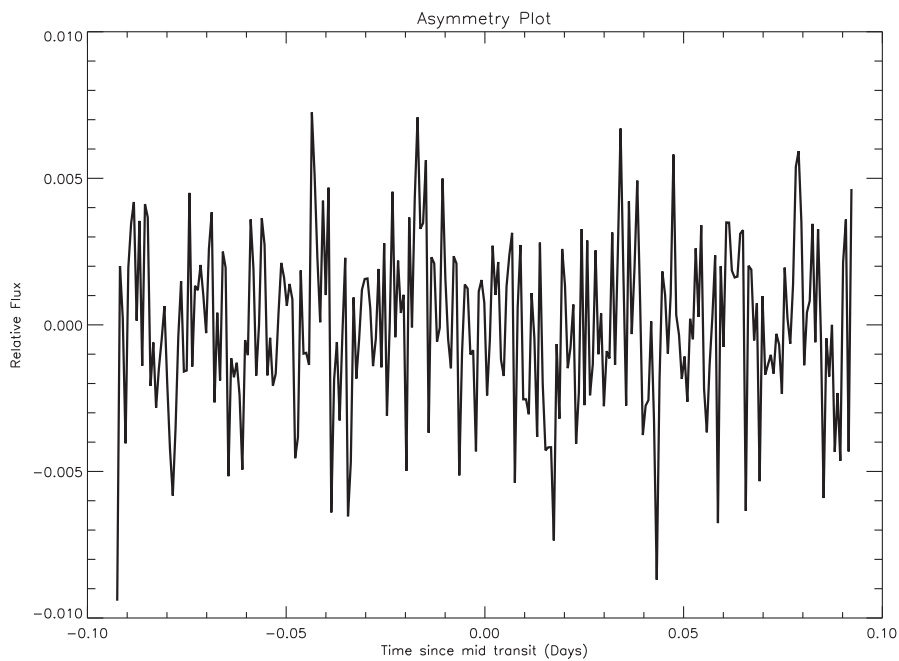

**Fig. 2.** The plot of the residuals from the transit subtracted by the mirror image of itself. Llama et al. (2011) found that the symmetry of a transit light curve is broken when a bow shock is introduced, resulting in an early near-UV ingress and a skewed shape of the transit ingress (the exact shape depends on the plasma temperature, shock geometry, and optical depth). We do not see an early transit ingress (assuming a reasonable planetary magnetic field) or an odd shaped transit ingress.

### 3. Light curve analysis

To find the best fit to the light curve we developed a modeling package called EXOMOP[4] and compared our results to another publicly available model, JKTEBOP[5] (Southworth et al., 2004a; Southworth et al., 2004b). EXOMOP uses the analytic equations of Mandel and Agol (2002) to generate a model transit, the Levenberg–Marquardt (LM) non-linear least squares minimization algorithm (Press et al., 1992) to find the best fit, the bootstrap Monte Carlo technique (Press et al., 1992) to calculate robust errors of the LM fitted parameters, a Differential Evolution Markov Chain Monte Carlo (DE-MCMC) (Braak, 2006) analysis to find the best fit and associated errors, and used both the residual permutation (rosary bead) method (Southworth, 2008) and time-averaging method (Pont et al., 2006) to access the importance of red noise in both fitting methods. JKTEBOP was originally developed from the EBOP program written for eclipsing binary star systems (Popper and Etzel, 1981a,b), and implements the Nelson–Davis–Etzel eclipsing binary model (Nelson and Davis, 1972). One advantage of JKTEBOP over EXOMOP is that it uses biaxial spheroids to model the object of interest (in our case a planet), allowing for departures from sphericity (Southworth, 2010). This feature of JKTEBOP is important in our analysis because if HAT-P-16b exhibits an early near-UV ingress, the planet would appear to be non-spherical (e.g., Llama et al., 2011, see Fig. 2). In addition, JKTEBOP uses the Levenberg–Marquadt Monte Carlo technique to compute errors (Press et al., 1992; Southworth, 2010; Hoyer et al., 2011). We only used JKTEBOP to search for any non-spherical asymmetries in the near-UV light curve (see Section 4.1)

We modeled the transit with the DE-MCMC using 15 chains and $2 \times 10^6$ links. The Gelman-Rubin statistic (Gelman and Rubin, 1992) was used to ensure chain convergence, as outlined in Ford (2006). We used the Metropolis-Hastings sampler and bayesian inference to characterize the uncertainties because it accounts for non-Gaussian errors and covariances between parameters. Our DE-MCMC model was derived from EXOFAST by Eastman et al. (2013).

---

[4] EXOMOP is available from http://sites.google.com/site/astrojaketurner/codes.
[5] http://www.astro.keele.ac.uk/jkt/codes/jktebop.html.

**Table 4**
Fixed EXOMOP parameters.

| Parameter | Symbol | Value |
|---|---|---|
| Period[1] (d) | $P$ | 2.77596 |
| Scaled semimajor axis[1] | $a/R_*$ | 7.17 |
| Inclination[1] (deg) | $i$ | 86.6 |
| Argument of periastron[1] (deg) | $\omega$ | 214 |
| Eccentricity[2] | $e$ | 0.034 |
| Linear limb darkening[3] | $\mu_1$ | 0.65720257 |
| Quadratic limb darkening[3] | $\mu_2$ | 0.17652937 |

[1] Reference Buchhave et al. (2010).
[2] Reference Husnoo et al. (2012).
[3] Reference Claret and Bloemen (2011).

**Table 5**
Derived EXOMOP parameters.

| Parameter | Symbol | Value |
|---|---|---|
| Planet-to-star radius ratio | $R_p/R_*$ | 0.10586 ± 0.00088 |
| Midtransit ime (BJD) | $T_c$ | 2456290.65779 ± 0.00049 |
| Transit duration (min) | $T_{14}$ | 183.42 ± 1.43 |
| RMS of white noise | $\sigma_{white}$ | 0.002344 ± 0.000024 |
| RMS of red noise | $\sigma_{red}$ | 0.0 ± 0.0 |

During the analysis, the time of mid-transit ($T_c$) and planet-to-star radius ($\frac{R_p}{R_*}$) were left as free parameters because we got non-gaussian distributions when fitting for $\frac{a}{R_*}$ and inclination ($i$) so we reverted to fix them at the discovery value. Eccentricity ($e$), argument of periastron ($\omega$), semi-major axis scaled to the size of the star ($\frac{a}{R_*}$), inclination ($i$), the quadratic limb darkening coefficients ($\mu_1$ and $\mu_2$), and the orbital period ($P_b$) of the planet were fixed to the values listed in Table 4. The linear($\mu_1$) and quadratic($\mu_2$) limb darkening coefficients for the Bessell-U filter were taken from Claret and Bloemen (2011) using the stellar parameters ($T_{eff} = 6158$ K, $\log g = 4.34$ (cgs), $[Fe/H] = 0.17$) from Buchhave et al. (2010). We used the fitted parameters from either the LM best-fit model or DE-MCMC best-fit model that produced the lowest scatter in the respective residuals (Transit–Best-fit model). The $\frac{R_p}{R_*}$ and $T_c$ parameters obtained from the EXOMOP analysis and the derived transit durations are summarized in Table 5.

To determine the error in the fitted parameters with the LM method we used the following bootstrap procedure. In step (1), we obtained the best-fit light curves and parameters from the LM non-linear least squares algorithm. In step (2), we multiplied the formal error bars for each data point in the light curve by random Gaussian noise with a standard deviation equal to the original error bars. In step (3), we added the error bars from step (2) to the data. In step (4), we repeated step (1) to find a new best-fit light curve. This process was repeated 10,000 times to avoid small-number statistics. When all iterations were finished the original best-fit value was subtracted from each fitted parameter from step (4) and a Gaussian function was fit to the distribution. The standard deviations of the distributions are the 1 sigma uncertainties in the fitted parameters.

In the residual permutation method the best-fit model is subtracted from the data, the residuals are then added to the data points. A new fit is found, and then the residuals are shifted again, with those at the end wrapped around to the start of the data. In this way, every new synthetic data set has the same bumps and wiggles as the actual data but only translated in time. Usually this process continues until the residuals have cycled back to where they originated. We updated this procedure by allowing for the error bars of the residuals to be taken into account. This is similar to step (2) in the bootstrap procedure described above. This process was repeated 10000 times. This procedure results in a distribution of fitted values for each parameter from which its uncertainty was estimated using the standard deviation of a Gaussian fit. For each fitted parameter we then defined $\beta_{res}$ (the scaling factor relative to white noise using the residual permutation method) as $\sigma_w/\sigma_{res}$, where $\sigma_w$ are the error bars derived from the bootstrap Monte Carlo technique or the DE-MCMC technique and the $\sigma_{res}$ are the error bars derived from the residual permutation method.

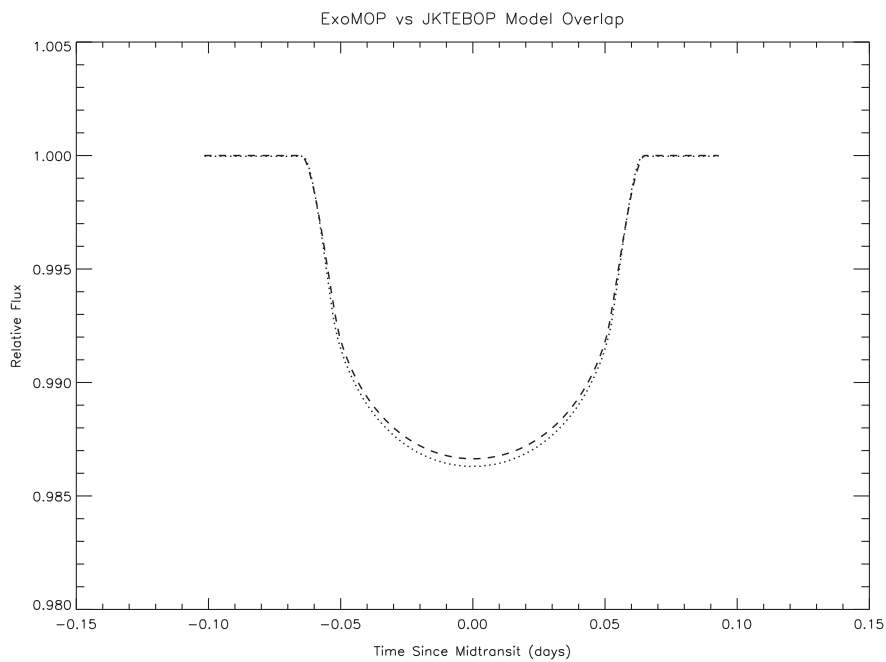

**Fig. 3.** The over lap between the best-fitting model obtained from EXOMOP (dashed line) and JKTEBOP (dotted line). We find a difference of 0.176 mmag between the two models. These results imply that the transit of HAT-P-16b is spherically symmetric.

We implemented the time-averaging method in a similar fashion to that done by Winn et al. (2008). For each light curve we found the best-fitting model and calculated $\sigma_{or}$, the standard deviation of the unbinned residuals between the observed and calculated fluxes. Next, the residuals were binned into bins of N points and we calculated the standard deviation, $\sigma_N$, of the binned residuals. In our analysis, N ranged from 1 to n/5, where n is the total number of data points in each respective transit. We then used a LM non-linear least squares minimization algorithm to find the RMS of red noise ($\sigma_{red}$) and the RMS of white noise ($\sigma_{white}$) using the following equation from Pont et al. (2006):

$$\sigma_N = \sqrt{\frac{\sigma_{white}^2}{N} + \sigma_{red}}. \quad (2)$$

The values for $\sigma_{white}$ and $\sigma_{red}$ for each transit can be found in Table 5. Using $\sigma_{white}$ and $\sigma_{red}$ we estimated $\beta_{tim}$, the scaling factor relative to white noise using the time-averaging method, with the following equation from Carter and Winn (2009):

$$\beta_{tim} = \sqrt{1 + \left(\frac{\sigma_{red}}{\sigma_{white}}\right)}. \quad (3)$$

To get the final error bars for the fitted parameters we multiplied $\sigma_w$ by the largest $\beta$ (either $\beta_{tim}$ or $\beta_{res}$) from either the residual permutation or time-averaging red noise calculations to account for underestimated error bars due to red noise (Carter et al., 2008). This was only done if the largest $\beta$ was greater than one. Finally, in cases where the reduced chi-squared ($\chi_r^2$) of the data to the best-fit model was found to be greater than unity we multiplied the error bars above by $\sqrt{\chi_r^2}$ to compensate for the underestimated observational errors (Bruntt et al., 2006; Southworth et al., 2007a). The final error bars can be found in Table 5 and our modeled light curve in Fig. 1.

We calculated the transit duration, $\tau_t$, of our transit model fit with the following equation from Carter et al. (2008):

$$\tau_t = t_{egress} - t_{ingress}, \quad (4)$$

where $t_{egress}$ is the best-fitting model time of egress, and $t_{ingress}$ is the best-fitting model time of ingress. To calculate the error in the $\tau_t$ we followed Carter et al. (2008) and set the error to $\sqrt{2}*$ cadence of our observation.

We also performed a similar analysis on our data utilizing the Transit Analysis Package[6] (TAP; Mandel and Agol, 2002; Carter and Winn, 2009; Gazak et al., 2012; Eastman et al., 2013) in order to check our EXOMOP results against a different MCMC based transit analysis package. We obtained results consistent with those from EXOMOP except TAP found red noise in our transit when EXOMOP found none. Turner et al. (2012b) found that TAP overestimates the red noise in light curves with only random Gaussian noise. Therefore, we decided to just use EXOMOP for our analysis.

### 3.1. Period determination

By combining our EXOMOP derived mid-transit times with a previously published mid-transit time of HAT-P-16b, we can refine the orbital period of the planet and this can be used in the future to search for any transit timing variations. The mid-transit time was transformed from JD, which is based on UTC time, into BJD, which is based on Barycentric Dynamical Time (TDB), using the online converter[7] by Eastman et al. (2010). Using our derived mid-transit time in Table 5 and the discovery value of $T_c$ = 2455027.65779 ± 0.00049 BJD$_{TDB}$, we derived an improved ephemeris for HAT-P-16b by performing a weighted linear least-squares analysis using the following equation:

$$T_c = BJD_{TDB}T_c(0) + P_bE, \quad (5)$$

where $P_b$ is the orbital period of HAT-P-16b and $E$ is the integer number of cycles after the discovery paper (Buchhave et al., 2010). From this derivation we obtained values of $T_{c(0)}$ = 2456290.65883 ± 0.00052 BJD$_{TDB}$ and $P_b = 2.7759690 \pm 0.0000013$ d. Our result is consistent with the discovery value with an error more precise by a factor of 2.

## 4. Discussion

### 4.1. Searching for asymmetry in the near-UV light curve

VJH11a predicted that HAT-P-16b should exhibit an early near-UV ingress. This effect has been observed in WASP-12b with five *Hubble Space Telescope* spectroscopic data points using the Cosmic Origins Spectrograph in the NUVA (253.9–258.0 nm) near-UV filter (FHF10), and multiple transits of TrES-3b have been observed while searching for this effect (Turner et al., 2013). FHF10 found that the near-UV light curve of WASP-12b started approximately 25–30 min earlier than its optical light curve, and also exhibited a near-UV $R_p/R_* \sim 0.1$ greater than in the optical (FHF10, see Fig. 2). Alternatively, several follow up observations by Haswell et al. (2012) used the same instrument on HST and concluded that there was too much variability in their transit to conclusively say their transit had an early ingress compared to FHF10. Additionally, Turner et al. (2013) found that the near-UV light curve of TrES-3b did not occur 5–11 min earlier as predicted by VJH11a. We do not observe such a large early ingress in the near-UV light curve or a significant $R_p/R_*$ difference in our transit of HAT-P-16b. As seen in Table 1, the planetary parameters ($M_b$, $R_b$, $P_b$, $a$, $R_*$, $M_*$, [Fe/H]) of WASP-12b and HAT-P-16b are different, and the WASP-12 $B_p/B_*$ value calculated from VJH11a is $\sim$11.8 times higher than the HAT-P-16b system. Using §1 we found a timing difference between the near-UV and optical ingress of 19–38 min assuming $B_*$ = 100 G and $B_p$ ranging from 8 to 40 G.

In an attempt to detect much smaller differences in the near-UV ingress (on the order of several minutes) and $R_p/R_*$ (on the order of $\sim$0.05) in the light curves of HAT-P-16b, we examined our data carefully and used different models to try to detect signs of transit asymmetry. Llama et al. (2011) presented models of the bow shock of WASP-12b, all with an apparent asymmetry between the two halves of the transit (Llama et al., 2011, see Fig. 2). Following this idea, we subtracted each light curve by the mirror image of itself about the EXOMOP calculated mid-transit time. We used this technique to find possible asymmetries on either half of the transit. From this test the RMS value was 2.89 mmag. Since this value was higher than the precision of the transit it would be possible to detect an asymmetry above the noise level, but our asymmetry plot in Fig. 2 has no underlying symmetry that would represent asymmetry in our transit. Fig. 3 also shows the EXOMOP best-fitting model over lapping the JKTEBOP best-fitting model. We compared our model to JKTEBOP because JKTEBOP assumes biaxial spheroidal symmetry and uses non-spheroidal shapes which will extend our ability to model an asymmetry. However, because it only uses biaxial spheroids it will not model a bow shock perfectly but it is a good preliminary step in detecting an asymmetry. By subtracting the EXOMOP best-fitting model from the JKTEBOP best-fitting model, we obtained a difference of 0.176 mmag between the two models that was below the precision of our light curve. From our result, it is clear that the near-UV transit did not display any non-spherical asymmetries since the JKTEBOP (which is capable of accounting for non-spherical asymmetries) and EXOMOP (which assumes the planet is spherical) models are nearly identical. This finding implies that HAT-P-16b is spherical in the near-UV. The results of this experiment indicate no asymmetries above the noise levels, and imply that our transit of HAT-P-16b is spherically symmetric. As shown in Fig. 1, we should have seen an early ingress of at least 19 min (assuming a reasonable $B_p$ and $B_*$) with the timing resolution and precision of our near-UV light curve. We cannot compare our near-UV transit shape with a predicted transit shape using VJH11a's model because Llama et al. (2011) found that the symmetry of a transit light curve is broken when a bow shock is introduced resulting in the shape of the transit ingress being skewed (the exact shape depends on the plasma temperature, shock geometry, and optical depth). We conclude that our near-UV light curve of HAT-P-16b is symmetrical and does not show any asymmetries.

---
[6] http://ifa.hawaii.edu/users/zgazak/IfA/TAP.html.
[7] http://astroutils.astronomy.ohio-state.edu/time/utc2bjd.html.

We observed a low significance characteristic in our light curve that could be explained by a very extended planetary magnetosphere. Vidotto et al. (2011b) predict that a double transit may be observed if the material absorbing near-UV radiation is concentrated in a small area around the bow shock (Vidotto et al., 2011b, see Fig. 9) in which it will transit before the planetary transit. The timing difference between the ingress of the supposed bow shock and our transit is ∼26 min with the bow shock beginning at a phase of −0.0305. Although if we project our error bars to greater than 2$\sigma$ this feature blends in with the rest of the noise in our transit, so we cannot say for certain if this feature comes from the planet's magnetic field or if it is an unknown systematic.

Furthermore, we can use existing measurements of the Rossiter–McLaughlin effect (McLaughlin, 1924; Rossiter, 1924; Winn, 2011) to assess whether HAT-P-16b is in the co-rotation radius of the host star and constrain possible bow shock variability. Vidotto et al. (2011b) predict that a bow shock will form ahead of the planet if the orbital velocity of the planet is greater than the rate at which the coronal material rotates. Moutou et al. (2011) constrained the spin-orbit alignment to be −10 ± 16° and a stellar rotational velocity of 3.9 ± 0.8 km s$^{-1}$ for HAT-P-16. The rotational velocity of HAT-P-16b is ∼160 km s$^{-1}$ by assuming a circular orbit with a radius equal to the semi-major axis and a period that we derived previously. As well, the azimuthal velocity of the co-rotating corona at the semi-major axis is ∼25 km s$^{-1}$. Clearly our approximation does not come close to the stellar or corona rotational velocity so we can conclude HAT-P-16b is subject to host a bow shock ahead of its orbital direction. Additionally, the constraint on the spin-orbit alignment could imply that the magnetic field locality of HAT-P-16 does not drastically change during the course of the orbit and that HAT-P-16b moves through coronal material of roughly similar density. However, HAT-P-16 is a moderately active star (subject to host turbulent fluctuations in the stellar wind, flaring, coronal mass ejections or variation due to a magnetic cycle) so it is possible that the inhomogeneity of coronal outflow and variability of the magnetic field will cause us to not detect a bow shock. Buchhave et al. (2010) measured the $R'_{HK}$ index of HAT-P-16 to be −4.86 which is the ratio between chromospheric activity in Ca II H and K emission to the total bolometric emission of the star (Noyes et al., 1984). This value is greater than that of the Sun (−4.96) suggesting that HAT-P-16 is considerably more active and certain to host active regions and therefore a non-uniform corona and inhomogeneous magnetic field. That being said it is likely the outflow of the star will be non-symmetric across the planetary orbit.

Even though we do not observe a timing difference in our near-UV light curve, it is still possible to calculate an upper limit to the magnetic field ratio of HAT-P-16b. The near-UV light curve of HAT-P-16b might contain a timing difference from an optical light curve, but it is just below the cadence sampled in our transits. The Nyquist–Shannon sampling theorem states that a signal has to be sampled at the Nyquist frequency (with a period equal to two sampling intervals) in order to fully detect the signal. Therefore, we use twice the near-UV light curve cadence for the input of the timing difference into Eq. (1). Any timing difference below this cannot be detected in our data set. Our cadence is much longer than the minimum timing difference predicted by VJH11a of 0.0042 s and shorter than the cadence achieved for the WASP-12b observations of 5733 s (FHF10). Assuming a timing difference of 120 s, we determine an upper limit on the magnetic field ratio of HAT-P-16b to be 0.0082$B_*$.

Our result implies that either the magnetic field of HAT-P-16b is abnormally low, or the VJH11a method to determine exoplanet magnetic fields cannot be applied to HAT-P-16b, or the effect proposed by VJH11a can only be observed with near-UV wavelengths not accessible from the ground (Turner et al., 2012a, 2013). The absence of a bow shock detection could be due to insufficient density of the shocked region to yield an observable signature. Or a weak stellar magnetic field, causing the static pressure balance between the planet and star to be incapable of hosting a bow shock (VJH11a, see Eq. (5)). In addition, our result may also suggest that the techniques outlined by VJH11a can only be used in narrow-band spectroscopy (as with the WASP-12b observations) and not broad-band photometry. However, a full radiative transfer analysis is needed to verify this possibility. Additionally, the spectral region (253.9–258.0 nm) covered by early near-UV observations of WASP-12b with the HST (FHF10) includes resonance lines from Na I, Al I, Sc II, Mn II, Fe I, Co I, and Mg I (Morton, 1991; Morton, 2000) and these species also have strong spectral lines in our U-band (303–417 nm; Sansonetti (2005)). Obtaining more observations of other exoplanets predicted by VJH11a to exhibit an early near-UV ingress will help distinguish between all the possibilities discussed here (Turner et al., 2013; Turner et al., 2012a, 2013).

Future observations with other telescopes capable of achieving a better near-UV cadence are needed to verify our conclusions, constrain VJH11a's techniques, and to search for bow shock temporal variations predicted by Vidotto et al. (2011c). We also advocate low-frequency radio emission and magnetic star-planet interaction observations of HAT-P-16b to further constrain its magnetic field and to supplement our findings.

### 4.2. Physical properties of the HAT-P-16 system

We used the results of our light curve modeling to calculate planetary and geometrical parameters of HAT-P-16b, including its mass, radius, density, surface gravity and Safronov number. Specifically, the EXOMOP values derived from our data analysis (Table 5) were used to derive the desired parameters.

We adopted the formula by Southworth et al. (2007b) to calculate the surface gravitational acceleration, $g_b$:

$$g_b = \frac{2\pi}{P_b}\left(\frac{a}{R_b}\right)^2 \frac{\sqrt{1-e^2}}{\sin i} K_*, \qquad (6)$$

where $K_*$ is the stellar velocity amplitude equal to 531 ± 2.8 m s$^{-1}$ (Buchhave et al., 2010).

The planetary mass, $M_b$, was calculated using the following equation derived from Seager (2011):

$$M_b = \left(\frac{\sqrt{1-e^2}}{28.4329}\right)\left(\frac{\gamma}{\sin i}\right)\left(\frac{P_b}{1yr}\right)^{1/3}\left(\frac{M_*}{M_\odot}\right)^{2/3} M_{jup}, \qquad (7)$$

where $\gamma$ is the radial velocity semi-amplitude equal to 531 ± 2.8 m s$^{-1}$ (Buchhave et al., 2010) and we used $M_* = 1.218^{+0.039}_{-0.039} M_\odot$ (Buchhave et al., 2010).

We calculated the Safronov number, $\Theta$, using the equation from Southworth (2010):

**Table 6**
Updated planetary parameters.

| Parameter | Symbol | Value |
|---|---|---|
| Period (d) | P | 2.7759690 ± 0.0000013 |
| Radius of planet ($R_{Jup}$) | $R_p$ | 1.274 ± 0.057 |
| Mass of planet ($M_{Jup}$) | $M_p$ | 4.194 ± 0.092 |
| Surface gravity (cgs) | log(g) | 3.81 ± 0.06 |
| Mean density (g/cm$^3$) | $\rho_p$ | 2.52 ± 0.34 |
| Safronov number | $\Theta$ | 0.223 ± 0.019 |

$$\Theta = \frac{M_b a}{M_* R_b}. \qquad (8)$$

The Safronov number is a measure of the ability of a planet to gravitationally scatter other bodies (Safronov, 1972). As defined by Hansen and Barman (2007), Class I hot Jupiters have $\Theta = 0.07 \pm 0.01$ and Class II have $\Theta = 0.04 \pm 0.01$.

Results of the $M_b$, $R_b$, the planetary density ($\rho_b$), $\log g_b$ and $\Theta$ from our analysis are summarized in Table 6. We used the values for the physical parameters of the HAT-P-16b system by Buchhave et al. (2010) and Husnoo et al. (2012) to derive the physical properties in Table 6. For all the planetary parameters, our results are within $1\sigma$ of published literature values.

We found a near-UV radius of $R_p = 1.274 \pm 0.057$ $R_{Jup}$, which is consistent with the near-IR radius of $R_p = 1.289 \pm 0.066$ $R_{Jup}$ Buchhave et al. (2010). We did not observe a wavelength dependence in the radius of HAT-P-16b through our analysis. Combining this result with previous near-IR studies, HAT-P-16b appears to have a wavelength-independent planetary radius in near-UV through near-IR wavelengths (Buchhave et al. (2010) used the HAT-6 and HAT-7 telescopes in Arizona and HAT-8 and HAT-9 telescopes in Hawaii to find a near-IR planetary radius of $1.289 \pm 0.066$ $R_{Jup}$) accessible from the ground, suggesting a nearly flat atmospheric spectrum in these wavelength bands. Cloud presence only in the upper atmosphere are consistent with our observation. Our result is consistent with other transiting exoplanet observations of a flat spectrum from visible to near-IR wavelengths on TrES-3b (a hot Jupiter; Turner et al., 2013), GJ 3470b (a hot Uranus; Biddle et al. (2013)) and GJ 1214b (a super-Earth; Bean et al., 2011). But inconsistent with other transiting exoplanet observations that do not have a flat spectrum from visible to near-IR wavelengths on HD 189733b (a hot Jupiter; Sing et al., 2011) and HD 209458b (a hot Jupiter; Sing et al., 2008). Nonetheless, an in-depth radiative transfer model needs to be done to fully understand what may be causing the flat atmospheric spectrum of HAT-P-16b.

## 5. Conclusions

We have investigated the primary transit of HAT-P-16b observed on December 29th, 2012 in the near-UV filter. In this study we derived a new set of planetary system parameters $M_b$ = 4.193 ± 0.092 $M_{Jup}$, $R_b$ = 1.274 ± 0.057 $R_{Jup}$, $\rho_b$ = 2.52 ± 0.34 (g/cm$^3$), $\log g_b$ = 3.81 ± 0.06, $\Theta$ = 0.223 ± 0.019. We find that HAT-P-16b's near-UV planetary radius of $R_p = 1.274 \pm 0.057$ $R_{Jup}$ is consistent within error of its near-IR radius of $R_p = 1.289 \pm 0.066$ $R_{Jup}$ (Buchhave et al., 2010). Comparing our results with the discovery paper values, HAT-P-16b appears to have a constant planetary radius in the near-UV and near-IR wavelengths. We did not detect an early ingress as proposed by VJH11a despite having a detectable timing difference in the range of 19–38 min ($B_*$ = 100 G and $B_p$ ranging from 8 to 40 G). Even though we did not detect an early ingress we are still able to find an upper limit to the magnetic field ratio of HAT-P-16b. Since we could not distinguish a timing difference below the Nyquist frequency, we used a timing difference of twice the maximum near-UV light curve cadence (120 s) to derive a range on the upper limit of HAT-P-16b's magnetic field ratio equal to $0.0082B_*$. Our value is consistent with the non-detection of the magnetic field of TrES-3b using the same method (Turner et al., 2013).

Based on this result, we advocate for follow-up studies on the magnetic field of HAT-P-16b using other detection methods such as magnetic star-planet interactions and radio emission, as well as observations with telescopes capable of achieving a better near-UV cadence to verify our findings, the techniques of VJH11a, and to search for a temporal bow shock variation predicted by Vidotto et al. (2011c). Our findings imply that the magnetic field of HAT-P-16b is abnormally low, or that the effect proposed by VJH11a can only be observed with near-UV wavelengths not accessible from the ground. To further constrain these possibilities, we encourage follow up observations of other exoplanets predicted by VJH11a to exhibit an early near-UV ingress. Lastly, an in-depth radiative transfer analysis is needed to determine whether VJH11a's techniques can only be used in narrow band spectroscopy and not broad-band photometry.

## Acknowledgments

We sincerely thank the University of Arizona Astronomy Club, the Steward Observatory TAC, the Steward Observatory telescope day crew, Dr. Phillip Pinto, Dr. Karin Ostrom, Charles Ostrom, the Associated Students of the University of Arizona, Lauren Biddle, Michael Berube and Robert Thompson for supporting this research. We would also like to thank the anonymous referee for their insightful comments during the publication process. This manuscript is much improved thanks to their comments.